\documentclass[%
 reprint,
 superscriptaddress,
 amsmath,amssymb,
 aps,
]{revtex4-2}

\usepackage{graphicx}
\usepackage{dcolumn}
\usepackage{bm}

\usepackage{xcolor}
\usepackage{upgreek}
\usepackage{comment}

\usepackage[sort&compress]{natbib}
\begin{document}

\preprint{APS/123-QED}

\title{Influence of Added Dye on Marangoni-driven Droplet Instability}

\author{Carola Seyfert}
 \affiliation{Physics of Fluids Group, Department of Science and Technology, Mesa+ Institute, Max Planck Center for Complex Fluid Dynamics and J. M. Burgers Centre for Fluid Dynamics, University of Twente, 7500 AE Enschede, The Netherlands}
 \email{c.seyfert@utwente.nl}
 
\author{Alvaro Marin}%
 \affiliation{Physics of Fluids Group, Department of Science and Technology, Mesa+ Institute, Max Planck Center  for Complex Fluid Dynamics and J. M. Burgers Centre for Fluid Dynamics, University of Twente, 7500 AE Enschede, The Netherlands}
 \email{a.marin@utwente.nl}

\date{\today}

\begin{abstract}

Multiphase flows are challenging systems to study, not only from a fundamental point of view, but also from a practical one due to the difficulties in visualizing phases with similar refraction indices. An additional challenge arises when the multiphase flow to be observed occurs in the sub-millimetric range. A common solution in experimental fluid dynamics is the addition of a color dye for contrast enhancement. However, added dyes can act as surface-active agents and significantly influence the observed phenomena. 
In the recently reported Marangoni bursting phenomenon, where a binary droplet on top of an oil bath is atomized in thousands of smaller droplets by Marangoni stresses, visualization is particularly challenging and the use of color dyes is naturally tempting. 
In this work, we quantify experimentally the significant influence of added methyl blue dye, as well as the obtained contrast enhancement. We find that the reaction of the system to such an additive is far from trivial and brings an opportunity to learn about such a complex phenomenon. 
Additionally, we propose a simple method to analyze the contrast enhancement as a function of dye concentration, with the aim of finding the minimum concentration of dye for successful imaging in similar systems.

\end{abstract}

\maketitle


\section{Introduction}
Many fluid phenomena across all scales are governed by fluid interfaces: emulsions, foams, liquid jets, liquid sheet break up, as well as droplets on surfaces; all being elemental features of nature and technology. 
One key characteristic of a fluid interface is its interfacial tension $\gamma_{A-B}$, which is defined for the combination of two fluids A and B forming the interface and can be interpreted as the required energy to create new interface $A-B$ per unit area. 
For the ubiquitous case of a liquid-air interface, from now on we will refer to it as the \emph{surface tension} of the liquid.

A liquid's surface tension is a crucial parameter, especially on very small length scales, for droplet spreading and wetting of surfaces \citep{Tanner1979, Bonn2009}, as well as droplet impact and splashing \citep{Aytouna2010, Josserand2016, Allen1975}, and liquid jet break-up \citep{Rayleigh1878, Eggers2008}. A prime understanding of surface tension effects is therefore extremely important for applications like atomization \cite{Eggers2008, Basaran2002drop}, inkjet printing \cite{Wijshoff2010dynamics}, film coating \cite{Scriven1988Coating} and many others.

Similar to other physical properties like density and viscosity, surface tension is dependent on, for example, temperature, and especially in the case of a multi-component liquid, the composition. Differences in surface tension within a liquid bulk lead to surface stresses, which induce a restoring flow in the bulk, commonly known as Marangoni flow \citep{Marangoni1871}. 
The surface tension gradient inducing the Marangoni flow can either occur due to temperature differences (thermal Marangoni flow), or due to concentration differences at the surface of different components. The latter may occur in multi-component liquids (e.g. ethanol and water) or in liquid solutions containing surface-active components (e.g. surfactants or salts). 
In either case it is important to note that the stress, which leads to the flow formation, originates at the fluid interface. As the flow restores the balance of surface tension, it is directed from the area of low surface tension towards the area of higher surface tensions.

\begin{figure*}
\centering
\includegraphics[width=1.85\columnwidth]{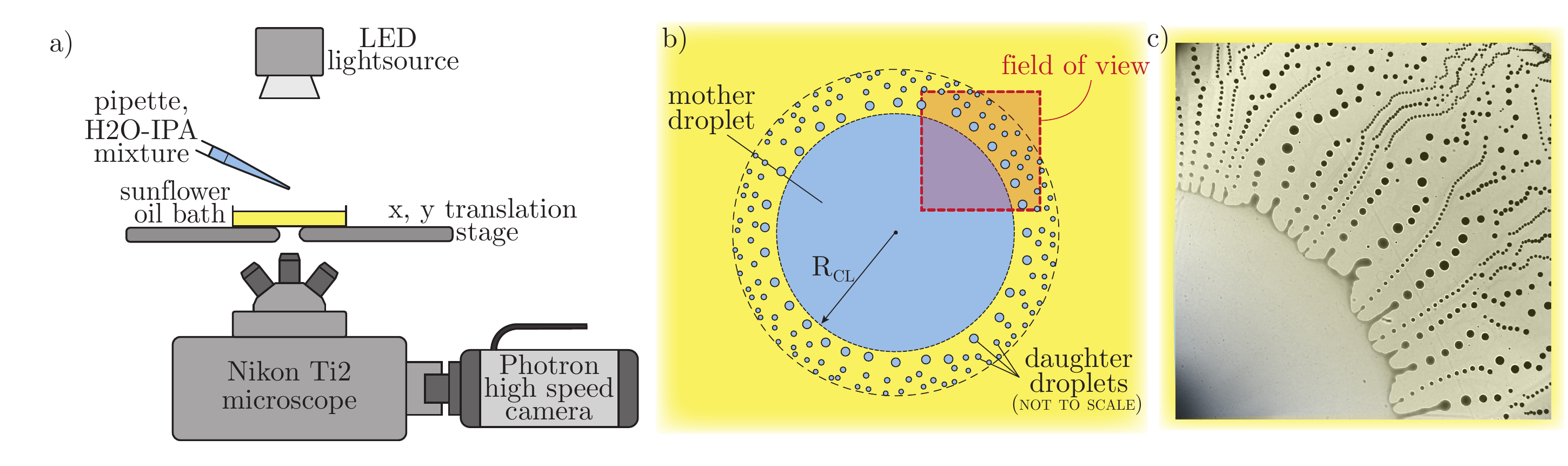}
\caption{
a) Schematic of the experimental set-up. 
b) Schematic of the experiment in an instance in time (not to scale). Illustrated are the mother (injected) droplet and the daughter droplets in light blue, on the sunflower oil bath in the background in yellow. The outer dashed circle denotes the outer bound of the experiment, i.e. the furthest the daughter droplets travel radially outwards. The inner dashed circle denotes the mother droplet radius, $R_{CL}$, which changes in time, as more and more alcohol evaporates. The unstable rim of the droplet and the ligament formation and droplet break-up are omitted in this schematic for sake of simplicity. The approximate position of the field of view of the microscope/camera is shown in red. We place a marker at the center on the bottom side of the Petri dish, to align our experiments such that the field of view position is comparable between different repetitions and experiments.
c) Partial image of the mother droplet and the emitted daughter droplets at high concentration of methyl blue dye.}
\label{fig:Setup-Marangoni}
\end{figure*}

In the context of droplets, both thermal and solutal Marangoni flows have been studied extensively for decades, unavoidably associated often with evaporation. The first qualitative observations of the thermal Marangoni flow in evaporating water droplets were made by Deegan et al. \cite{Deegan1997,Deegan2000contact}, and soon after quantitatively modeled by \citet{Hu2005}. Solutal Marangoni flows in this context have also been studied extensively, either to study their role in the spreading and retraction of multi-component droplets \cite{Guena2007,Williams2020}, or their role in the droplet's shape \citep{Tsoumpas2015, Guena2007,Karpitschka2017}, or their competition with gravitational forces \cite{Diddens2021competing}, or their role in particle deposition when driven by surfactants (or antisurfactants) \cite{Truskett2003,Still2012,Sempels2013biosurfactant,Marin2016surfactant,Marin2019salt,Bruning2020}. 

The phenomenon commonly known as the ``tears of wine'' shares all these previously mentioned elements. As correctly explain by \citet{Thomson1855} in 1855, and quantitatively analyzed much later by \citet{Fournier1992}, a solutal Marangoni flow initiated by non-homogeneous evaporation lifts a thin film of liquid along the wine glass, which thickens and becomes the well-known tears of wine that we observe in such joyful occasions. The tears-of-wine effect requires non-homogeneous evaporation, strong Marangoni flow and a freely moving contact line. When such conditions are fulfilled, a similar phenomenon can be observed for a multi-component solution droplet placed on a hydrophilic substrate, as shown by \citet{Mouat2020}. 

Combining these elements but placing the multi-component droplet on a liquid bath, \citet{Keiser2017} reported the so-called Marangoni bursting phenomenon, in which the ``tear formation'' is dramatically enhanced and the deposited droplet bursts into a myriad of daughter droplets. The particular case that \citet{Keiser2017} studied was generated with a droplet containing a mixture of water and alcohol, deposited on a sunflower oil bath. A similar effect was obtained by making use of a drop of an insoluble surfactant solution on a liquid bath by \citet{Wodlei2018}.

The beauty of the experiment of \citet{Keiser2017} lies in its apparent simplicity, since most of the components can be fairly easily obtained out of a laboratory environment, and it serves as a beautiful and illustrative experiment to use in outreach activities \citep{Durey2018Bursting}. One challenge of this otherwise simple and highly reproducible experiment is its visualization. Alcohol, water and sunflower oil share very similar refractive indices, resulting in extremely low contrast images of the phenomenon. A straightforward solution to this commonly encountered problem (and worsened by the small scale features) is to add a color dye to the system and thus enhance the imaging contrast. 
For many problems adding dye is the only option for successful imaging. However, it is often unclear whether the addition of dye also alters the intrinsic characteristics of the experiment itself. 
As recently described by \citet{Manikantan2020} in their perspective article, molecules do not need to have an amphiphilic structure to induce an effect at the liquid's interface and to be considered surface-active. Typically, we assume that in low concentrations, most dyes will have a negligible effect on the analyzed dynamics. It is tempting to make use of interfacial/surface tensions in (semi-)equilibrium to confirm the lack of influence of additives, but such measurements do not need to correlate well with experiments involving interfacial out-of-equilibrium processes. As we will show in our results, this is precisely the case of the Marangoni bursting phenomenon.

In this work, we present experimental results on the influence of added methyl blue dye on the Marangoni bursting phenomenon as an example of a small-scale, instability-driven fluid dynamics experiment. Methyl blue is an acid dye commonly used in histology and in fabric coloring. Most common food dyes also fall into the category of acid dyes. 
Methyl blue is a commonly used dye in fluid physics experiments and has also been used in other studies on Marangoni bursting \citep{Keiser2017, Hasegawa2021}, as well as in drop impact experiments \cite{Tuan2018splash,yang2019nonsphere}, droplet microfluidics \cite{liu2007droplet,Garstecki2019droplet}, interfacial instabilities \cite{figueiredo1996morphological} or visualization of (bulk) flow instabilities \cite{willmarth1964steady}. 
In the following, we first explain the methodology, then quantify the dye influence, explain the observed effects and finally aim to quantify the imaging enhancement due to the added dye. 

\begin{figure*}
\centering
\includegraphics[width=1.9\columnwidth]{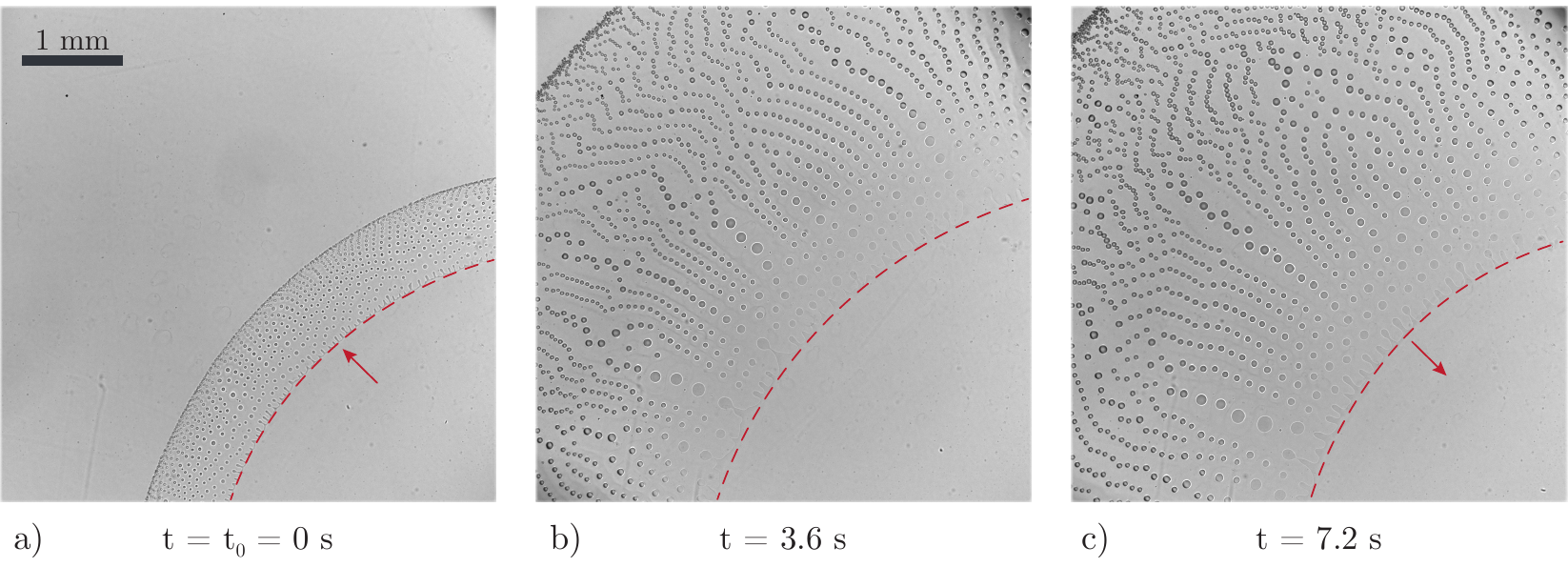}
\caption{Selected images of a typical experiment. Note the timestamps in every panel. The scale bar in panel a) also applies to the other two panels. In each panel, the mother droplet contact line is indicated with a dashed red line. In panel a) $\rm{t=t_{0}}$ as it is the first frame of the recording which is in focus. The daughter droplets as well as the contact line of the mother droplet are both spreading outwards (indicated with the red arrow). In panels b) and c) we can see the outer bounds of the experiment in the top left corners. The daughter droplets and the pinch-off sites along the unstable rim are visible. The contact line of the mother droplet has come to a halt in panel b) and is receding in panel c).}
\label{fig:SnapshotsExp-100xMB}
\end{figure*}

\section{Materials and Methods}
\label{sec:MatMet}

\subsection{Experimental Methods}
\label{sec:methods}

The images for this study are taken with a Photron Nova S12 high speed camera, at 2000 frames per second. The camera is connected to a Nikon Ti2 inverted microscope. We use a lens with a 4x magnification factor. The acquired 16bit images have resolutions of 4.8 $\upmu \rm{m}$ per pixel, and the field of view measures $\rm{\approx~24~mm^{2}}$. For a schematic of the setup and of the experiment, see Figure \ref{fig:Setup-Marangoni}.

In the following, we describe a typical experiment: we place a plastic Petri dish, containing sunflower oil up to a height of $\rm{\approx~1.5mm}$ (constant for all experiments), on the translation stage of the microscope. As the experiment is much bigger than the image dimensions, we achieve a reproducible alignment by placing a marker at the center of the bottom of each Petri dish, and placing the marker just outside the corner of our field of view. 
With an LED light source fixed above the stage, backlighting is achieved through the sample.
We inject an initial droplet volume of $\rm{1.5~\upmu l}$, which we call from now on ``mother droplet'', on top of the oil bath with a micro-pipette. The same initial mother droplet volume is kept constant throughout the study. The choice for that particular volume has been done according to imaging constraints, i.e. a compromise between a sufficiently large field of view and high resolution, which should not present any issue since the phenomenon seems to be independent of the mother droplet volume \citep{Keiser2017} at these scales. In Figure \ref{fig:SnapshotsExp-100xMB} we show three images captured during a typical experiment.
The droplet first spreads radially outwards and the contact line enters the field of view, shedding daughter droplets, which travel further radially outwards on top of the oil bath and leave the field of view. As the mother droplet evaporates, the contact line halts and then starts to recede, moving inwards through the field of view, all the while still shedding daughter droplets. Our recordings end when the mother droplet has left the field of view and only daughter droplets are visible.

Every experimental configuration (i.e. concentrations of dye) is repeated at least three times with fresh oil baths. 
After the image acquisition, we analyse various features of the experiment, such as the instability wavelength, the mother droplet radius, and droplet size distributions, as well as the imaging contrast. 
To find the mother droplet radius, we fit a circle along the contact line of the droplet, that is visible in the field of view.
The instability wavelength is obtained through two different methods: i) the arclength between two pinch-off sites, determined as the mother droplet radius multiplied by the angular difference between pinch-off sites; ii) the linear distance between two pinch-off sites. Since the wavelength is found to be much smaller than the mother droplet radius (at least three orders of magnitude), the linear approach is justified. Both methods to determine the wavelength yield indistinguishable results, with differences well below the measurement error. The reported wavelengths on each frame corresponds to the average of wavelengths within each time instance.

We define daughter droplet size distributions based on a region of interest close to the contact line. Individual daughter droplets travel radially outwards from the contact line, but often stay within the field of view for the majority of the experiment. They shrink in size as they evaporate further. To gain insight on the instability behavior and how it influences the droplet sizes, we include only ``freshly'' pinched-off daughter droplets in the size distributions.

The imaging contrast we report on in Section \ref{sec:ContrastEnhancement} is determined with two different methods: i) based on the average image intensity of a region of interest; ii) based on a distance transformation of the pixel intensity with the contact line as reference.

\begin{figure*}
\centering
\includegraphics[width=1.9\columnwidth]{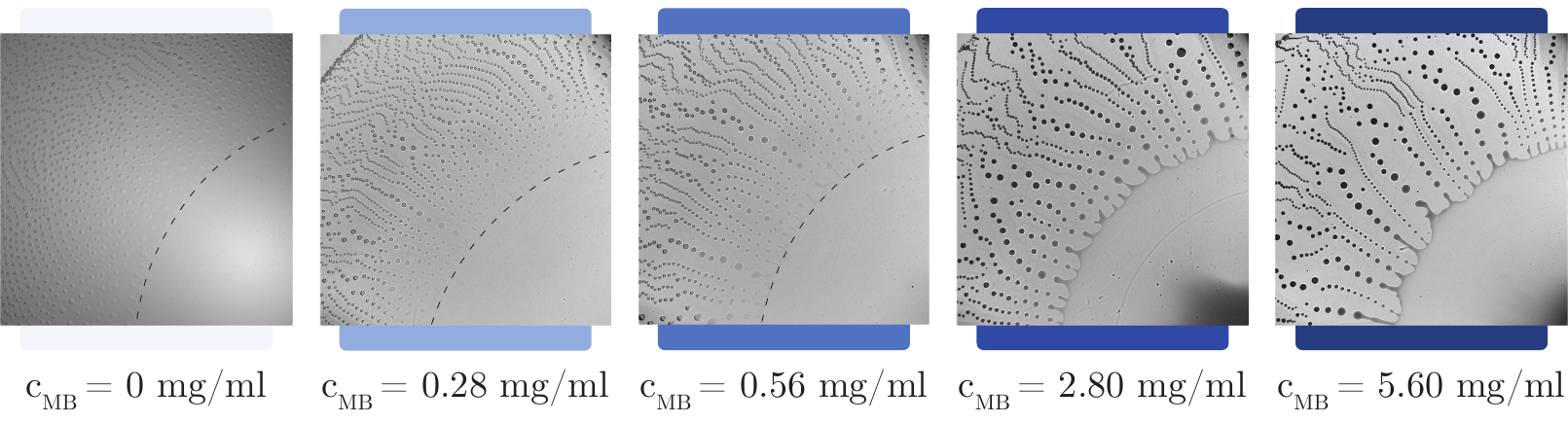}
\caption{Example images of experiments with varying methyl blue concentrations, at comparable time instances of experiments. From left to right, as $\rm{c_{MB}}$ increases, we see i) ligament formation at the contact line and a change in droplet pinch-off, as well as ii) a significant increase in imaging contrast. Note that, for better comparison of the original imaging contrast, we choose to show all images in this figure without any postprocessing. As a visual aid, we added dashed lines at the contact line positions in the first three images for which the imaging contrast is low. The images capture 4.9~mm in width by 4.9~mm in height. The background in each panel is consistent with the chosen concentration color map of later plots.}
\label{fig:DyeConcExampleImages}
\end{figure*}

\subsection{Liquids and Dye}
\label{sec:materials}
The three liquid phases necessary for the Marangoni bursting effect are water, alcohol and sunflower oil. For our experiments, we use purified water [from aMilli-Q\textsuperscript{\circledR}~IQ 7000 Water Purification System], standard laboratory isopropylalcohol (IPA) [2-Propanol, technical grade], and commercial food supply sunflower oil. The concentrations of IPA in mixtures are, if not stated otherwise, given as a volume fraction. 

We use a water soluble acid dye, methyl blue (MB) [Methyl Blue from Alfa Aesar]. We characterize the final concentration of dye $\rm{c_{MB}}$ as mass of dissolved dye per volume of water-IPA mixture, in units of mg/ml. The acid dye methyl blue is readily dissolved in water, but less so in IPA. Therefore, we first mix a certain mass of dye in the water phase, before adding the IPA to reach the final concentration of methyl blue in the mixture. We vary the dye concentration systematically and use the saturation concentration of methyl blue in water, $\rm{c_{MB, sat}}=70~$mg/ml, as a reference. For concentrations too close to $\rm{c_{MB, sat}}$, we observe precipitation and formation of solid agglomerates. Therefore, we choose the concentrations of dye accordingly, such that no precipitation of dye agglomerates is visible in the images we obtain from the experiments. In this study, we span a methyl blue concentration range of $\rm{c_{MB}}~=~0.28~\textup{--}~5.60~$mg/ml.

The influence of the added dye is expected to manifest most prominently in the interfacial tensions involved in the phenomenon; in particular at the interface between the mixture drop and air, $\rm{\gamma_{da}}$, and at the interface between the mixture drop and the oil bath,  $\rm{\gamma_{do}}$. During our experiments, the IPA concentration within the drop changes due to evaporation. The presence of methyl blue introduces another potentially surface-active component on top of the IPA concentration gradient.

Measurements of the interfacial tensions as a function of the MB concentration have been performed using the pendant drop method and can be found in the Appendix. From these measurements we can conclude that MB indeed has a surface-active effect, especially on the water/oil interfacial tension, in a lower degree in the presence of IPA, and almost undetectable at the surface tension (drop-air interface). Such measurements should be taken with caution, since they are performed in quasi-static conditions, unlike the conditions far from equilibrium found in the experiments.

\section{Marangoni bursting in the presence of added dye}
\label{sec:ResultsDye}

In the following we describe the observed effects of methyl blue on the Marangoni bursting phenomenon. For a more detailed description of the phenomenon itself, we refer to the previous paper by \citet{Keiser2017}.
They described in particular the effect of the initial IPA concentration in the mother droplet, $\rm{\upphi_{IPA,0}}$: the higher $\rm{\upphi_{IPA,0}}$, the smaller the instability wavelength. This also leads to smaller daughter droplets for high $\rm{\upphi_{IPA,0}}$. 
\citet{Keiser2017} reported a critical IPA mass fraction $\rm{\upphi_{IPA,0}\simeq 0.41}$, below which the bursting phenomenon does not occur any more. On the opposite end, for $\rm{\upphi_{IPA,0}}>0.85$ the thin outer regions of the droplet tend to rupture during the spreading, instead of progressing into a thickened outer rim. 
In our experiments, we observe the phenomenon for a similar range of IPA volume fractions of $\rm{\upphi_{IPA,0}}\simeq 0.45$ to $\rm{\upphi_{IPA,0}}>0.85$. Such a small difference in concentration range can very likely be caused by the different commercial sunflower oil employed in our experiments.

In order to explore the influence of added dye on the Marangoni bursting effect, we choose an initial IPA concentration $\rm{\upphi_{IPA,0}}=0.6$, as for this intermediate concentration the phenomenon is observed to be highly reproducible. In the following, the IPA concentration of any experiment is assumed to be 60~vol\%, unless stated otherwise.
The highest concentration of methyl blue we report on in this study, $\rm{c_{MB}}=5.60~$mg/ml, corresponds to $\rm{14~g/l}$ in the water phase (which makes up 40~vol\% of the mixture). This concentration is therefore a fivefold dilution of the saturation concentation $\rm{c_{MB, sat}}=70~g/l$.

In Figure \ref{fig:DyeConcExampleImages} we show selected images for different dye concentrations. The figure shows how the instability at the mother droplet's contact line becomes increasingly more erratic as we add more dye to the initial mix. The droplet rim is much more corrugated, with some of the ligaments at the pinch-off sites becoming longer and thicker, before the daughter droplets pinch off. 
In the reference case with no added dye ($\rm{c_{MB}}=0~$mg/ml), though challenging to see, the contact line is smoother and the daughter droplets break off of the rim with no significant ligament formation.

\begin{figure}
\centering
\includegraphics[width=0.98\columnwidth]{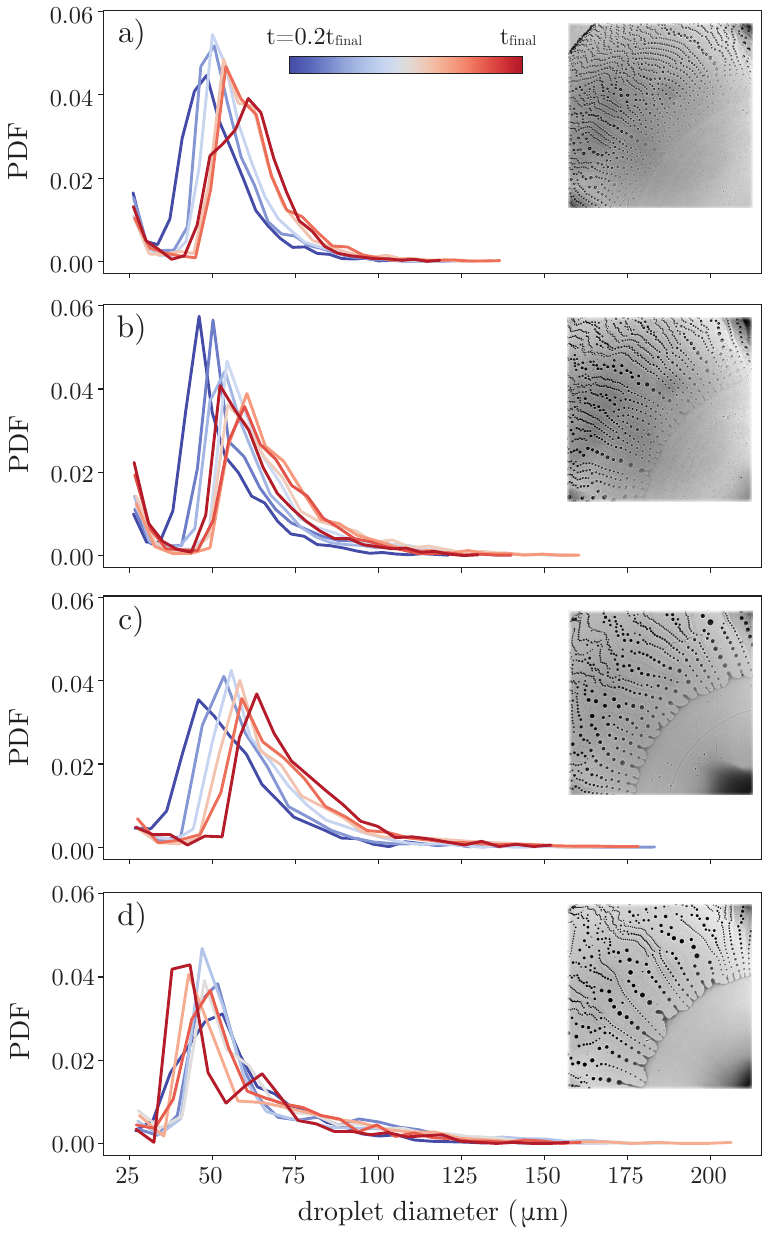}
\caption{
Probability density functions of daughter droplet diameters for different dye concentrations. 
From panel a) to d), $\rm{c_{MB}}$ corresponds to 0.28~mg/ml, 0.56~mg/ml, 2.8~mg/ml, and 5.6~mg/ml.
The different colors of individual distribution plots indicate the time (see the colormap in panel (a), valid for all panels). 
}
\label{fig:DropletSizeDistributions}
\end{figure}

\subsection{Daughter droplet size distributions}

\begin{figure*}
\centering
\includegraphics[width=1.9\columnwidth]{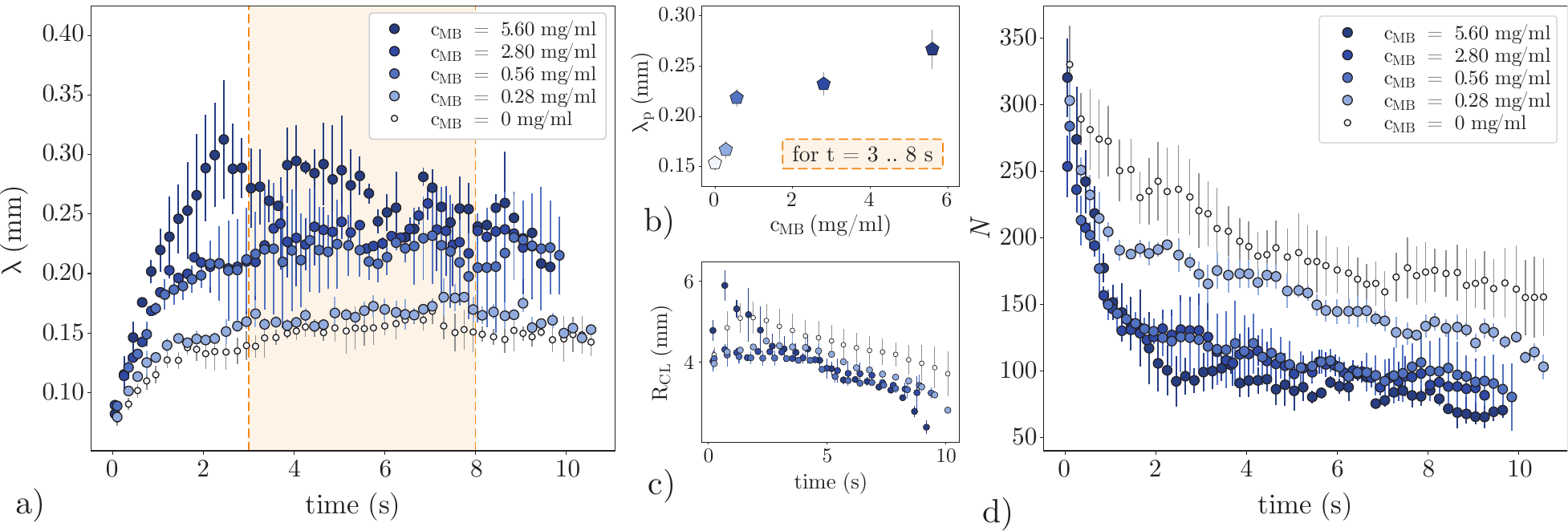}
\caption{
a) Instability wavelength $\lambda$ at the unstable droplet rim over time, for different MB concentrations. Plots show average of at least two experiments, with the standard deviation as error bar. The increase in wavelength in the beginning results from the outward spreading contact line. Added dye increases the instability wavelength. Between t=3~s and t=8~s (orange shaded plot area) we calculate a \emph{plateau wavelength $\lambda_{p}$} as an average of data points.
b) Plateau wavelength $\lambda_{p}$ for varying MB concentration.
c) Contact line radius $R_{CL}$ over time. The same color code applies as in the other panels. $R_{CL}$ for all dye concentration is smaller than the reference case ($\rm{c_{MB}}$=0).
d) Instability wavelength scaled with the circumference, $2\pi R_{CL}$, corresponds to the estimated total number of pinch-off sites $N$ around the mother droplet. Added dye leads to a significant drop in $N$.}
\label{fig:ResultsDye}
\end{figure*}

Figure \ref{fig:DropletSizeDistributions} shows daughter droplet size distributions for four different concentrations of dye from $\rm{c_{MB}}$= 0.28 to 5.6 mg/mL. The images from the reference case ($\rm{c_{MB}}=0~$mg/ml) are of such low contrast, that it is impossible to extract reliable data for droplet sizes.
In the very beginning of the process, as the contact line spreads out rapidly, the wavelength is still growing (see Figure \ref{fig:ResultsDye}) and the droplet size increases dramatically, before it reaches a quasi-steady plateau. Therefore, we choose to start plotting droplet size distributions at $t = 0.2\times t_{final}$ when the contact line motion has slowed down from the initial spreading to reach $R_{CL, max}$. We define $t_{final}$ as the frame when the contact line has left the field of view and only daughter droplets are visible.
In order to exclude contamination and dirt particles in the size distributions, a lower limit for the droplet detection is set at 24 $\upmu$m in diameter for all distributions, well below the main droplet curves. Note that below this detection limit one could also find satellite droplets that might form during droplet pinch off, which are consequently not included in the statistics.

As one can see in the size distributions, those from $\rm{c_{MB}}=0.28~$mg/ml to $\rm{c_{MB}}=2.80~$mg/ml (the top three panels) share similar trends. First, one can note that the droplets appear to grow with time, with the late distributions (red curves) being shifted to the right compared to the initial distributions (blue curves). In most cases, the distributions also become broader with time.
Second, while the general shape and the position of the maximum is comparable between $\rm{c_{MB}}=0.28~$mg/ml to $\rm{c_{MB}}=2.80~$mg/ml, the distributions become broader towards the larger droplet size as the dye concentration increases. The probability of finding significantly larger droplets increases as we increase the dye concentration. This is also manifested in the maximum droplet size of the distribution, which stretches to the right side of the plot as we increase the initial dye concentration. Such bigger droplets pinch off from long, thick ligaments, which are completely absent in experiments without dye. 

Unexpectedly, the distributions for the highest dye concentration $\rm{c_{MB}}=5.60~$mg/ml follow different trends. The size distributions shift towards smaller droplet sizes as time evolves, especially during late times of the experiment. At the latest times, the size distribution shows a maximum located at even smaller droplet sizes than for the lowest dye concentration $\rm{c_{MB}}=0.28~$mg/ml. Nevertheless, the tail of the distribution on the larger droplet size does follow the same trend as seen before, showing certain probability of finding much larger droplets. Although notoriously visible in the images, note that these large droplets account only for a small fraction in the total number of droplets. Volume distributions cannot be computed with our current measurements since the precise droplet shape is a priori unknown. 

In conclusion, we observe that the influence of added dye on the daughter droplet size distributions is twofold: i) adding more dye leads to the formation of bigger droplets, as a consequence of the presence of thicker and longer ligaments and ii) for the highest dye concentrations, a wider range of droplets sizes coexist, and the temporal trend of the size distribution is more erratic.

\subsection{Instability wavelength and rim's radius}
The daughter droplet size distribution is a direct consequence of the edge instabilities developing at the mother droplet's rim.  Figure \ref{fig:DyeConcExampleImages} shows how ligaments at certain pinch-off sites become thicker, longer, and more stable as the dye concentration increases, which, in turn, is reflected in the instability wavelength. 

Figure \ref{fig:ResultsDye}a) shows the instability wavelength $\lambda$, for different concentrations of methyl blue. The wavelength increases at the initial instances for all experiments while the droplet spreads. Eventually the wavelength stabilizes at a plateau value, once the contact line has reached its maximum diameter and starts to recede again. In panel b) of Figure \ref{fig:ResultsDye} we plot the \emph{plateau wavelength} $\lambda_{p}$ over the methyl blue concentration. To compute $\lambda_{p}$ we calculate the average of $\lambda$ between t=3~s and t=8~s. 

Despite the difficulties analysing experiments without dye due to the minimal imaging contrast, we find that $\lambda_{p}\simeq 0.15~$ mm, in the same order of magnitude reported earlier in literature \citep{Keiser2017, Hasegawa2021}. For $\rm{c_{MB}}=2.80~$mg/ml the plateau wavelength is $\lambda_{p}\simeq 0.23~$ mm, i.e. a 50\% increase with respect to the case without dye. The highest dye concentration probed in this study,  $\rm{c_{MB}}=5.60~$mg/ml, yields even higher values for the wavelength, however the fluctuations in wavelength also grow proportionally with the amount of added dye (also see Figure \ref{fig:DyeConcExampleImages}), in agreement with the erratic behavior observed for the droplet size distributions.

There is a clear correlation between the instability wavelength and the mother droplet radius, as can be seen from the increase in wavelength at the beginning of each experiment. In order to compensate for such dependence, we calculate the number of pinch-off sites $N$ as the circumference of the droplet (i.e. a function of the droplet radius) over the wavelength

\begin{equation}
N~=~\frac{2 \pi R_{CL}}{\lambda}~.
\label{equ:numberLigaments}
\end{equation}

Note that in our current setup we do not have optical access to the full droplet perimeter and therefore $N$ in this case is an indirect average estimation of the number of pinch-off sites. 
$N$ decreases significantly over time, as the mother droplet decreases in size and the circumference shrinks. Figure \ref{fig:ResultsDye} shows that adding dye to the experiment reduces $N$ by up to a factor 2 (panel d)), which corresponds to an increase in wavelength $\lambda$ (panel a)) and a decrease in contact line radius $R_{CL}$ (panel c)).

\subsection{Discussion}

Similar trends as those observed here by the addition of MB dye are observed for a decrease of initial IPA concentration in the mother droplet: the lower $\rm{\upphi_{IPA,0}}$, the longer the wavelength \citep{Keiser2017}, which yields larger daughter droplets. This effect is caused through the interfacial tension gradient which builds up at the mother droplet's interfaces during the experiment. In the following we will first recall the mechanisms behind the reference case with no added dye ($\rm{c_{MB}}=0$ mg/ml), and then discuss the implications when the added dye is present.

At all time instances, the volume fraction of IPA at the rim of the mother droplet is considered to be the critical concentration of $\rm{\upphi_{IPA,crit}\approx 0.4}$. We assume that the center of the droplet retains an IPA concentration $\rm{\upphi_{IPA}} \approx \rm{\upphi_{IPA,0}}$. In the following, an effective surface tension of the droplet is considered as the sum of the interfacial tensions $\gamma = \rm{\gamma_{da} + \gamma_{do}}$, following \citet{Keiser2017}. Such an effective surface tension $\gamma$ is dependent on the alcohol concentration and will always be lowest at the droplet center, as the alcohol concentration will be highest there for all instances. In comparison, in vicinity of the droplet's rim we expect to find $\rm{\upphi_{IPA}} \approx \upphi_{IPA,crit}$, corresponding to the highest surface tension in the mother droplet.

The surface tension gradient $\frac{\Delta\gamma}{\Delta x}$ building up between center and rim of the droplet during evaporation -- the main driving force in the system  -- is therefore steeper for higher initial IPA concentrations:

\begin{equation}
\Delta \gamma~=~\gamma_{crit}~-~\gamma_{0}~,
\label{equ:DeltaGamma}
\end{equation}

with $\gamma_{crit}$ the droplet surface tension in regions with $\rm{\upphi_{IPA}} \approx \upphi_{IPA,crit}$, and $\gamma_{0}$ the droplet surface tension in regions with $\rm{\upphi_{IPA}} \approx\rm{\upphi_{IPA,0}}$, with $\gamma_{crit}>\gamma_{0}$ at all times. Note that our only control parameter is $\rm{\upphi_{IPA,0}}$. 

Consequently, an increase in $\rm{\upphi_{IPA,0}}$ is followed by an increase in $\Delta \gamma$, and a stronger Marangoni stress at the interfaces, which leads to shorter instability wavelengths $\lambda$.

Interestingly, the addition of dye yields the inverse effect (figure \ref{fig:ResultsDye}): a higher $\rm{c_{MB}}$ (which typically involves lower effective surface tension) leads to longer $\lambda$. We will therefore now analyze the influence of $\rm{c_{MB}}$ on the surface tension gradient in the mother droplet.

The mild decrease of interfacial tensions by the addition of methyl blue dye (see our measurements shown in the appendix of this paper) seems to be responsible for the main observed effects: an increase of average droplet size and a longer wavelength.  In a multiphase droplet undergoing Marangoni bursting, the dye would therefore fulfill a somewhat similar role as IPA, except for being nonvolatile, compared to the highly volatile IPA. This is a crucial difference between both additives: while the IPA concentration decreases as we approach the rim, we can safely expect a larger concentration of non-volatile dye at the droplet's rim. The main reason for such a dye enrichment at the rim would be the evaporation of IPA -- most prominently -- and water.

These facts lead to various implications which can be connected to our experimental observations and from which we can learn something about the response of this complex system to additives:
\begin{enumerate} 
\item A secondary concentration gradient of methyl blue can superimpose on and even counteract the primary 
concentration gradient of IPA in the droplet. While $\rm{\upphi_{IPA}}$ at the rim is lower than in the droplet center, $\rm{c_{MB}}$ at the rim is likely higher than in the center. A high methyl blue concentration at the edge could lower the critical surface tension $\gamma_{crit}$, leading to a reduced overall surface tension gradient. 
\item A high local $\rm{c_{MB}}$ at the edge of the droplet can locally lower the surface tension, and hence favor the formation of long, stable ligaments. Our experimental observations show these long and stable ligaments for high dye concentrations, whereas no ligaments form in a case with no added dye.
\item Additionally, the presence of stagnation points at the rim between ligaments could cause additional dye concentration gradients in the angular direction along the perimeter, which would explain the multimodal ligament size distribution for the largest initial MB concentrations and the wider droplet size distributions.
\end{enumerate}

So far we have shown that the presence of methyl blue dye significantly affects the experiment. Nonetheless, it is important to stress the ``positive'' effects of dye, i.e. the acquired contrast enhancement, as we will do in the following section.

\section{Quantification of Contrast Enhancement}
\label{sec:ContrastEnhancement}

Dyes are generally used to enhance contrast to improve image analysis. This has to be balanced with eventual side effects that might have an impact on the observed phenomenon. In order to find the best balance, it is crucial to quantify these effects. 

As described previously in our particular case, methyl blue dye has an influence in the Marangoni bursting effect proportionally to the dye concentration.
In the following we quantify the contrast enhancement for varying concentrations of dye. Imaging contrast can be broadly defined as the difference between the maximum and minimum values of an image's histogram. To compute this difference, we use two different methods, both based on a region of interest (ROI) around the contact line of the droplet. Since the original frames are slightly vignetted in the corners (see for example Figure \ref{fig:SnapshotsExp-100xMB} and Figure \ref{fig:DyeConcExampleImages}), and as we are mostly interested in the center of the image and the droplet edge, we crop the image to a ROI around the droplet rim. 

In a direct and averaging approach to calculate the imaging contrast, we compute the intensity difference as $I_{0.95}-I_{0.05}$ for the ROI; to avoid extreme outliers, we defer from using the maximum and minimum values of the histogram, and opt for the 5\% and 95\% values instead. Normalizing this difference by the bit depth of 16bit, we get a single value for the \emph{overall imaging contrast} per frame. Furthermore, we calculate the average of 200 frames (arbitrarily chosen), where the droplet contact line stays within the ROI. 

\begin{figure*}
\centering
\includegraphics[width=1.98\columnwidth]{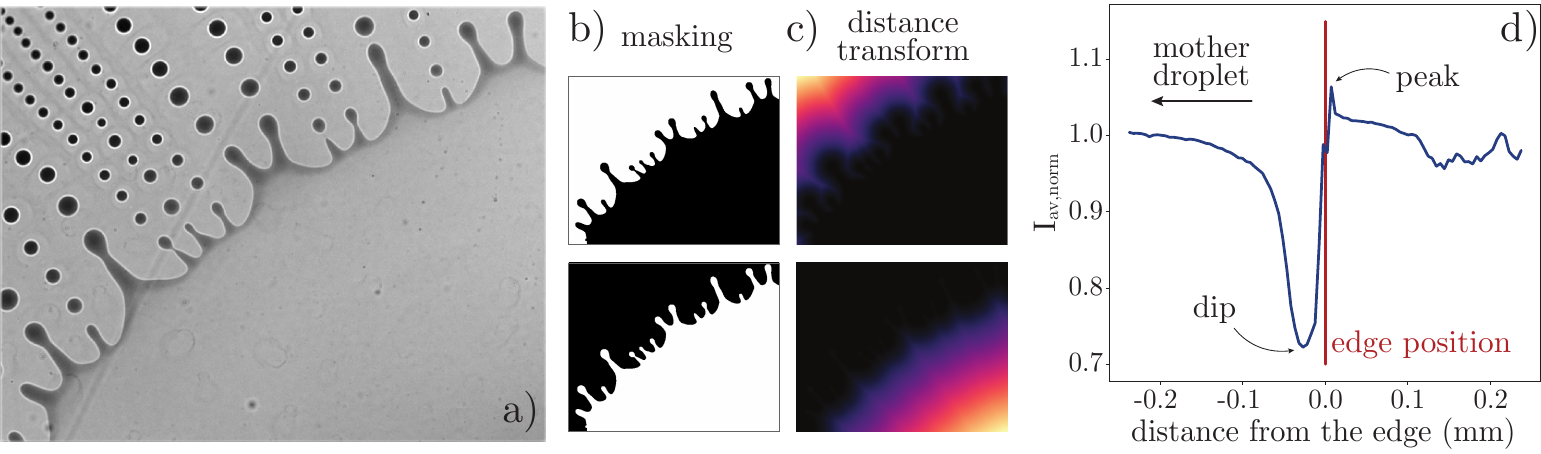}
\caption{Example for imaging contrast quantification with a distance transformation based on the droplet edge. 
An example image is shown in panel a) of an experiment with $\rm{\rm{c_{MB}}=5.60~mg/ml}$. 
In panel b) we show the two masks created from the cropped original, in order to get the distance transforms for inside the mother droplet, as well as outside; see panel c). 
In panel d) we plot the average image intensity $\mathrm{I_{av,norm}}$ over the distance from the edge, with the edge position denoted with a vertical red line. The intensity is averaged over all pixels with the same distance from the edge (sign sensitive). For better comparison with other dye concentrations, we normalize the averaged intensity with the mean intensity inside the mother droplet.}
\label{fig:ContrastDistTrans}
\end{figure*}

With a second, more detailed approach we find the pixel intensity as a function of the distance from the droplet edge. For this approach, we first detect the edge of the droplet, and then perform a distance transformation on a masking image (see Figure \ref{fig:ContrastDistTrans}). For all pixels within the image, that are located at the same (sign sensitive) distance from the edge, we calculate an average intensity. We use example images for different dye concentrations. Furthermore, we normalize the pixel intensity for each distance with the mean average intensity inside the mother droplet (as this intensity is in itself dependent on the dye concentration). A good measure of the contract enhancement \emph{across} the droplet edge can be the difference between the dip and the peak in intensity, right before and right after the edge (see panel d) in Figure \ref{fig:ContrastDistTrans}).

\begin{figure}
\centering
\includegraphics[width=\columnwidth]{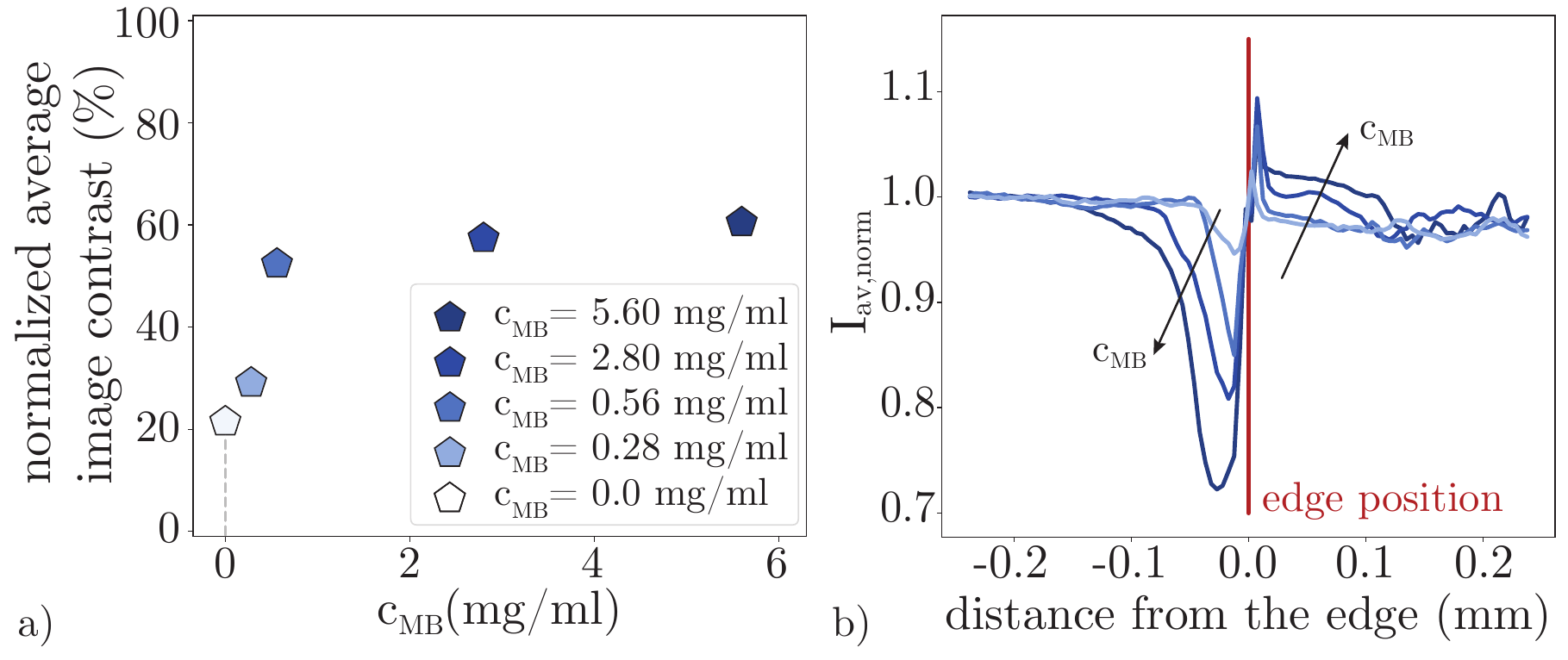}
\caption{a) Average overall imaging contrast as a function of dye concentration. We normalize the overall imaging contrast found as intensity difference with the bit depth of 16 bit. The leftmost data point represents a reference experiment with no dye. The contrast enhancement with dye leads to a rise in overall intensity to $\rm{\sim 60\%}$. After reaching this plateau value, adding more dye does not yield higher overall intensity any more. b) Contrast quantification based on distance transformation for different dye concentrations. The colors of the plots correspond to the legend in panel (a). Note that it was not possible to compute this parameter in the absence of dye due to the lack of contrast for the masking image. We observe a significant contrast enhancement across the droplet edge with increasing dye concentration.}
\label{fig:ContrastQuanti}
\end{figure}

Both quantification methods yield information about the contrast enhancement due to the added dye, see Figure \ref{fig:ContrastQuanti}. While the averaging approach gives an overall contrast of an image, the distance transformation offers detailed information on how well the droplet edge can be detected. As can be seen in panel a), Figure \ref{fig:ContrastQuanti}, the overall intensity increases significantly as soon as dye is added to the system, but then seems to reach a plateau value for dye concentrations at around 0.7~mg/ml. Adding more dye to the mixture, does not significantly enhance the imaging contrast further. 

However, if we take a closer look at the contrast enhancement across the contact line, as shown in panel b) of Figure \ref{fig:ContrastQuanti}, we can see that higher dye concentrations in the system yield better enhancement, without the same plateau effect as for the overall intensity.

For the experiments in this study, even a minimal amount of added dye ($\rm{c_{MB}}=0.28~$mg/ml, which yields almost no contrast enhancement) affects the bursting phenomenon. The wavelength increases only minimally, but since the contact line radius also decreases compared to the reference case ($\rm{c_{MB}}$=0 mg/mL), the number of pinch-off sites already decreases significantly (see Figure \ref{fig:ResultsDye}). For higher dye concentrations, which yield a sufficient contrast enhancement, the effects on the bursting phenomenon are even bigger.

\section{Conclusions}
\label{sec:Conclusions}

In this work we show that the influence of adding dye for imaging contrast enhancement needs to be evaluated carefully. Since some phenomena do not provide enough imaging contrast to be analyzed properly without dye, it is crucial to balance the contrast enhancement with the undesirable effects of the dye on the experiment itself.

In this particular case of the Marangoni bursting phenomenon, the instability wavelength and the daughter droplet sizes and numbers are influenced significantly by the amount of added methyl blue dye. This is probably due to a decrease in interfacial tensions. However, surprising effects are found at the highest dye concentration such as a multimodal daughter droplet size distribution, which cannot be explained by the decrease of stationary interfacial tensions alone. Although perhaps trivial in some specific  fields, we find it important to reiterate here that stationary interfacial tension measurements are not enough to discard the influence on an additive on a phenomenon driven by interfacial instabilities. This message is very clearly shown in our results.

Although mode advanced experimental measurements would help to better understand the processes taking place (e.g. fluid flow and film thickness measurements), the ``hidden''\cite{Manikantan2020} nature of surface-active material requires the use of numerical simulations for a full understanding of such effects, which are beyond the scope of this paper.

Additionally, we quantify the imaging contrast enhancement with an averaging intensity measurement, as well as with a more detail-oriented distance transformation approach, with a characteristic image feature as reference (in this case the droplet edge). In this study we observe a sufficient increase in imaging contrast for dye concentrations $\rm{c_{MB}}\gtrsim0.7~$mg/ml, however we also show that even for the lowest probed dye concentration $\rm{c_{MB}}=0.28~$mg/ml (which provided no significant contrast improvement), the characteristics of the experiment change. Similar experimental protocols should always be followed when employing color dyes for visualizing multiphase flow phenomena, which we hope to have exemplified properly here.

\begin{acknowledgments}

The authors acknowledge financial support from the European Research Council, Starting Grant No. 678573 NanoPacks.
The authors thank Maziyar Jalaal, Detlef Lohse, Jacco Snoeijer and Christian Diddens for fruitful and inspiring discussions, 
as well as Sander G. Huisman for his most valuable input on the contrast quantification.

\end{acknowledgments}

\section*{Appendix}
\label{app:IFT}

\begin{figure*}
\centering
\includegraphics[width=1.9\columnwidth]{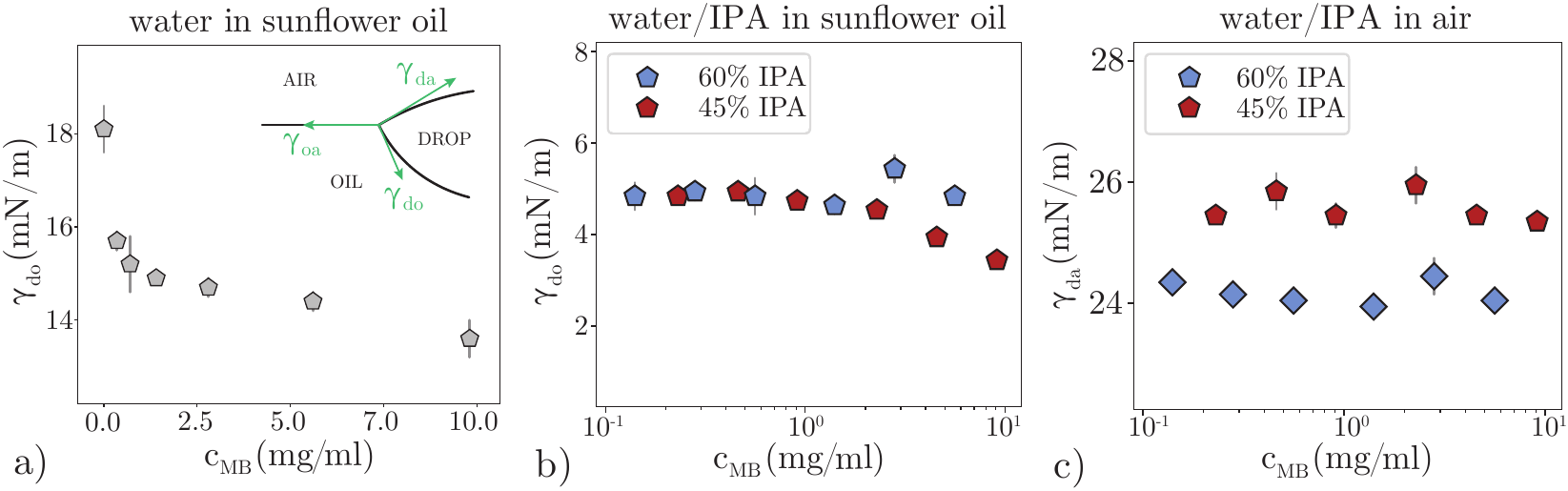}
\caption{
a) Interfacial tension between water and sunflower oil, for an increasing concentration of methyl blue. 
b) interfacial tension for water and IPA mixtures (with the initial mother drop concentration in this work, 60vol\% IPA, as well as the critical concentration of IPA, present at the unstable rim, of 45vol\% IPA), for an increasing concentration of methyl blue.
c) Interfacial tension for water and IPA mixtures (60vol\% and 45vol\% IPA) for an increasing methyl blue concentration.
}
\label{fig:IFT}
\end{figure*}

We measure interfacial tensions $\gamma$ using the pendant drop method, in a commercial contact angle measurement setup (DataPhysics, OCA 15). Each experiment is repeated several times and the value shown corresponds to the average of at least three measurements. The standard deviation of such measurements set is used as the measurement error shown in Figure \ref{fig:IFT}. 
Throughout the literature, one can find a wide range of values for the interfacial tension between sunflower oil and water \citep{Dragosavac2008, Bhatluri2015, Bai2018, Cong2020}. We observe some discrepancies within the reported values, comparable to our own measurements. This is reasonable given the variability of commercial sunflower oils used in these cited studies. Different processing and refining methods lead to different physical properties, such as the interfacial tensions, of interest for us. 

We observe a decrease in interfacial tension $\rm{\gamma_{do}}$ for pure water droplets in sunflower oil with an increasing methyl blue dye concentration. The change in $\rm{\upgamma_{do}}$ measures up to 30\% between the pure water case and a solution of methyl blue in water.


\end{document}